\documentclass{sigchi}

\usepackage{balance}  
\usepackage{graphics} 
\usepackage{times}    
\usepackage{helvet}
\usepackage{url}      
\usepackage[hide]{chato-notes} 
\usepackage{caption,subcaption,float,booktabs,enumitem}

\makeatletter
\def\url@leostyle{%
  \@ifundefined{selectfont}{\def\UrlFont{\sf}}{\def\UrlFont{\small\bf\ttfamily}}}
\makeatother
\def\pprw{8.5in}
\def\pprh{11in}

\usepackage[pdftex]{hyperref}
\hypersetup{
pdftitle={SIGCHI Conference Proceedings Format},
pdfauthor={LaTeX},
pdfkeywords={SIGCHI, proceedings, archival format},
bookmarksnumbered,
pdfstartview={FitH},
colorlinks,
citecolor=black,
filecolor=black,
linkcolor=black,
urlcolor=black,
breaklinks=true,
}

\begin{document}

\title{Building Bridges into the Unknown:\\Personalizing Connections to Little-known Countries\\}

\toappear{\scriptsize Permission to make digital or hard copies of all or part of this work for personal or classroom use is granted without fee provided that copies are not made or distributed for profit or commercial advantage and that copies bear this notice and the full citation on the first page. Copyrights for components of this work owned by others than ACM must be honored. Abstracting with credit is permitted. To copy otherwise, or republish, to post on servers or to redistribute to lists, requires prior specific permission and/or a fee. Request permissions from permissions@acm.org. \\
{\emph{CHI 2015}}, April 18--23, 2015, Seoul, Republic of Korea. \\
Copyright \copyright~2015 ACM 978-1-4503-3145-6/15/04\ ...\$15.00. \\
http://dx.doi.org/10.1145/2702123.2702152}
\clubpenalty=10000 
\widowpenalty = 10000

\numberofauthors{1}  
\author{
  \alignauthor Yelena Mejova, Javier Borge-Holthoefer, Ingmar Weber\\
    \affaddr{Qatar Computing Research Institute}\\
    \email{\{ymejova,jborge,iweber\}@qf.org.qa}
    \vspace{0.6cm}
}

\maketitle

\begin{abstract}
How are you related to Malawi? Do recent events on the Comoros effect you in any subtle way? Who in your extended social network is in Croatia? We seldom ask ourselves these questions, yet a ``long tail'' of content beyond our everyday knowledge is waiting to be explored. In this work we propose a recommendation task of \textbf{creating interest in little-known content by building personalized ``bridges'' to users}. 
We consider an example task of interesting users in little-known countries, and propose a system which aggregates a user's Twitter profile, network, and tweets to create an interest model, which is then matched to a library of knowledge about the countries. We perform a user study of 69 participants and conduct 11 in-depth interviews in order to evaluate the efficacy of the proposed approach and gather qualitative insight into the effect of multi-faceted use of Twitter on the perception of the bridges. 
We find the increase in interest concerning little-known content to greatly depend on the pre-existing disposition to it. Additionally, we discover a set of vital properties good bridges must possess, including recency, novelty, emotiveness, and a proper selection of language. Using the proposed approach we aim to harvest the ``invisible connections" to make explicit the idea of a ``small world'' where even a faraway country is more closely connected to you than you might have imagined.
\end{abstract}

\keywords{Recommendation; Serendipity; Personalization; Filter bubble}

\category{H.3.3}{Information Search and Retrieval}{}
\category{H.3.4}{Systems and Software}{}


\section{Introduction}
\label{sec:introduction}

%
%
%

Since Travers and Milgram \cite{travers69} provided experimental evidence on the {\em effective size} of the social space -- quantified at six hops on average -- our small world has shrunk even further. Time scales have virtually vanished and geographical distance has become irrelevant through the advent of the Internet and the Web 2.0: we now live in an ultra-small world \cite{cohen2003scale}. Indeed, breaking news from any point on Earth pop up, within minutes, in our collection of connected devices, while we whimsically interact with peers around the globe.

In practice, this idealized vision is hindered due to psychological, social and technological constraints. Our world is {\em potentially} an ultra-small one, as long as any piece of information is just a few steps away from a given source; and yet a bubble hampers our capacity to reach out. Possible culprits are our limited attention span \cite{weng2012competition}, a preference towards similar \cite{mcpherson2001birds} and nearby acquaintances \cite{takhteyev2012geography}, and the filtering layers that mediate our online experience \cite{pariserfilter}: news feeds and social connections that are chosen to get more of what we like and agree with, and less of what challenges our beliefs or is simply not of interest. Remarkably, however, not everything within the bubble is agreeable and exciting; and --presumably-- not everything outside of it is offensive or indifferent to our preferences.

The aim of this work is then to propose a break-through exercise, an attempt to discover (or create) interest beyond the ``comfort zone''. Unlike the standard recommender system setting, wherein the recommendation algorithm is like a salesman that first tries to understand your needs and interests and then recommends a product that you are likely to buy, in our setting, the unfortunate salesman only has a single item that you may be initially not interested in. 


Our methodology, then, aims to provide personalized {\em bridges} to connect social media users to any content.

\begin{quote}
A \emph{bridge} between a user $u$ and a content $c$ is a bit of information that increases $u$'s interests in $c$. We are particularly focusing on \emph{personalized bridges} where the bit of information makes an explicit link between $u$'s interests, origin or social network and content $c$. 
\end{quote}

\begin{figure*}[ht]
\centering
\includegraphics[width=0.8\textwidth]{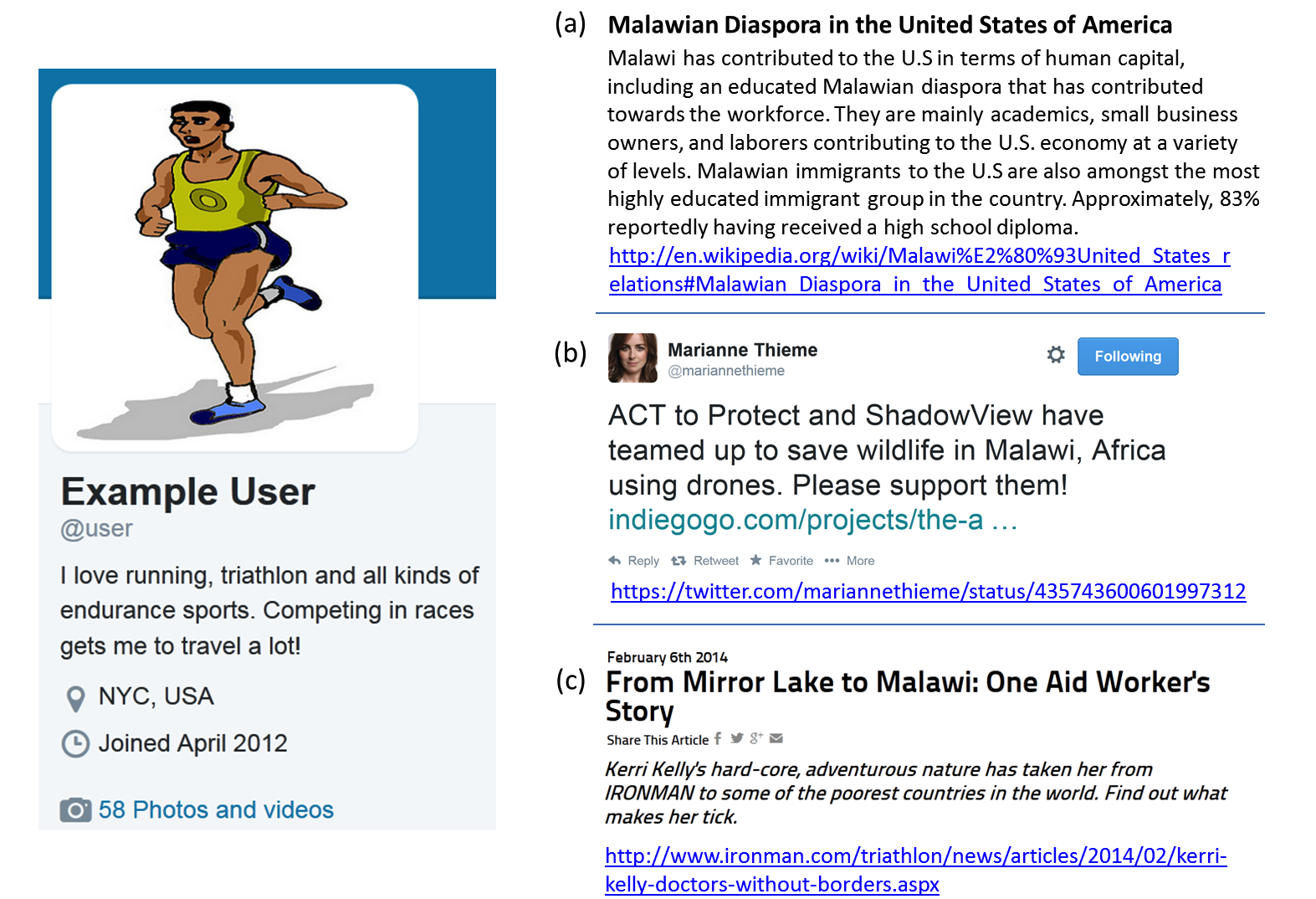}
\vspace{0.2cm}
\caption{An illustrative example of our approach. On the left, the profile for @user, a fictitous Twitter user, is shown. On the right, three types of ``bridges'' linking this user to the country Malawi are shown. Bridge (a) uses the fact that the user has specified ``NYC, USA'' as their location. Bridge (b) is a relevant tweet from the user's friendship network on Twitter. Bridge (c) exploits the user's interest in triathlon. The goal is to increase the user's interest in Malawi through one or more of such bridges.}
\label{fig:example}
\end{figure*}

In this paper, we consider a case study of relating users to various countries. To find such bridges, we build a user interest model (via their Twitter account) and a country model (using several knowledge sources), and discover user-country connections that may fill the ``gap'' between the two. Note that such gap might originally already be a small one, e.g.\ between a user and their home country, but the challenge lies in distant connections. ``Distant'' should here be understood in its widest meaning: literally distant places, but also culturally remote, economically unrelated or with poor media coverage. Figure~\ref{fig:example} shows a fictitious example where bridges are built between a Twitter user from the U.S.\ who is interested in triathlons and the South-East African country Malawi. Possible bridges may include Wikipedia articles (a), tweets from their friends (b), or news stories (c).



To evaluate bridges utilizing a variety of sources, we conduct a user study, generating personalized bridges for $69$ Twitter users, and conduct further interviews with $11$ participants. As a result, we provide insight into the user's treatment of little-known content, information coming from their social network, and the multi-faceted behavior our system encounters.

Although our current system is targetting countries, the insights we produce empirically are generalizable to many other kinds of target content. The issues of personal closeness to the members of one's personal network, the attractiveness of media, recency, and novelty are all applicable to building connections via user interests and network.  Beyond our motivating application of broadening a user's news diet, our approach can contribute to the area of Data Science for social good, as it may raise awareness for ``forgotten places'' during humanitarian crises or natural disasters. 
It could become one potential tool to help ``rewire'' the world to create true cosmopolitans \cite{zuckerman13rewire}.

\begin{quote}
\emph{It's hardly possible to overstate the value, in the present state of human improvement, of placing human beings in contact with other persons dissimilar to themselves, and with modes of thought and action unlike those with which they are familiar.}
\end{quote}
\vspace{-0.2cm}
\hfill -- John Stuart Mill


\section{Related Work}
\label{sec:relatedwork}
For a helpful contextualization of our work, we focus on the select research avenues within the domain of recommendation systems which include context-aware approaches, those which incorporate social network information, or propose to explain their recommendations. Below we spotlight only those which share some features that are relevant to our proposal. 

First, we highlight works deviating from similarity-based recommendation, which aims to recommend items for which similar people expressed some interest. In this less-populated niche, the idea is to break through the filter bubble either relying on serendipity (Auralist \cite{zhang2012auralist}, for example, attempts to introduce diversity into one's musical playlist), or targeting purposefully contrarian standpoints (\cite{graells2013data} explores visualization of a law debate). We specially focus our attention to the {\em Terra Incognita} project (\url{https://terra-incognita.co/}) which has significant overlap with this proposal in that (i) it has a central geographical ingredient; and (ii) it attempts to relate users with locations (cities in this case) through the mediation of news articles. Also, the project is related to international news coverage biases \cite{kwak2014first,zuckerman04nieman}, a variable that affects our work as well, as we discuss later. It lacks however any sort of personalization and does not attempt to build bridges.

Besides the aforementioned {\em Terra Incognita}, other works have paid attention to the recommendation of places exploring location-based social networks \cite{symeonidis2013geosocialrec}. More application-oriented, we find a rather large literature which is focused solely in touristic recommendation \cite{moreno14,ricci2002travel} -- attractions, monuments, travel packages, etc. Some of these, mainly in the domain of ``smart systems'' and Artificial Intelligence, share with the current work the ``proactive recommendation'' trait (in the sense of unasked for), but fall unequivocally in the similarity-based recommendation set.

Third, there is a large body of work on the construction of a user model. Abel {\em et al.}\ \cite{abel2011analyzing} rely not only on internal Twitter activity (namely, user profile and authored tweets), but also on the content of the news that the user visits to build up personalization. From this point of view, it is already a scheme that transcends the single user and reaches out to other contents for which the user has shown interest. Closer to the present work, De Francisci Morales {\em et al.}\ \cite{de2012chatter} build a user model which relies on the ego-network, i.e.\ information from the user and from their immediate (more influential) neighbors. They also provide a ``news model'' -- as we build a ``country model'' -- which is to be confronted with the user to seek plausible recommendations. More generally, social network information has been modeled in the context of recommendation by Walter {\em et al.}\ \cite{walter2008model}, who propose that users ``leverage their social network to reach information and make use of trust relationships to filter information''. Also, graph-based approaches have quantified the relationships among both the users and the items by weighing the graph edges during a random walk traversal \cite{konstas2009social}. In the traditional recommendation setup these approaches incorporate the social network information into their item scoring algorithms. In this work, we instead use the network information explicitly to motivate (that is, ``bridge'') the recommendation to the user, and provide detailed user feedback on this strategy.

We also address the privacy concerns over the use of personal information in recommendation. In exploring potential data sources other than Twitter, we discuss the viewpoints of potential users on expanding data collection to their Facebook profiles. Although a study in 2005 has found people were ``quite oblivious'' about their privacy settings on Facebook \cite{gross2005information}, the later publicity and interface changes have increased privacy-seeking behavior in its users \cite{stutzman2013silent}. Our findings support this phenomenon and further illustrate the complicated division between professional and personal online personae.  

Finally, it is possible to identify some works which devote their efforts to provide explanations, which enhance the user's experience and understanding of the recommendation. Such explanations may come in the form of illustration and graphics \cite{herlocker2000explaining} or ``data portraits'' and ``organic visualization design'' \cite{graells2013data}; tags \cite{vig2009tagsplanations}; or more complex, phrased explanations \cite{blanco2012you}. Our study focuses on the particular challenges of using Twitter as a data source for links and explanations, specifically providing a list of desirable attributes (motivated by extensive interview process) and methodological approaches to achieve them.



\section{Building Bridges}
\label{sec:buildingbridges}

The nature of our experiment requires a collection of data associated with (i) the user and (ii) the country to which we are trying to uncover an interest. Below we describe the process of collection, selection, and cleaning of the data for both, and the bridges we build using it.

\subsection{Twitter User Data}

In order to personalize the bridges, we first need to understand the user's interests, which we attempt using public Twitter data. We collect data using the Twitter APIs, and more precisely, the \texttt{twitter} Python library\footnote{\url{https://code.google.com/p/python-twitter/}}. First, we collect the user profile data, including the handle, screen name, self-assigned location string, description, and profile image URL. We then proceed to collect the $3,200$ latest tweets by the user. We use this data in order to build a model of the user's interests by applying the following steps:

\begin{itemize}
\item The tweet text and the profile description are lower-cased and cleaned of special characters, URLs and handles. The text is then tokenized to result in 1-grams (single words), which can be counted to generate an initial language model.
\item These 1-grams are then filtered using several stopword lists, including a comprehensive list of English-language words and the top 500 Twitter terms obtained from a general Twitter sample (using Decahose).
\item A part of speech tagger is then applied to the remaining terms (using Stanford NLP toolkit\footnote{\url{http://nlp.stanford.edu/software/tagger.shtml}}) and only nouns (or plural nouns) are selected. 
\item The tweets are also cleaned and tokenized to produce 2- and 3-grams (2 and 3 word potential phrases). 
\item The 1-, 2-, and 3-grams are then merged such that the frequency of lower-$n$ grams is reduced by that of higher-$n$ grams which include them (so the count for ``social'' does not include the count for ``social media''). In this process, we favor the higher-$n$ grams (as they are likely to express a more precise concept), and select all candidates which pass an empirically-set frequency threshold of $3$. 
\item Finally, the language model of the profile is added to the resulting list with an assumption that it contains concepts user chose as most important in describing themself on the website.
\end{itemize}

The resulting terms are often places, such as cities and countries in which users live (such as \emph{texas} or \emph{france}), concepts associated with their professional interests (\emph{university}, \emph{engineering}), and other interests (\emph{hogwarts}, \emph{cat}, \emph{salsa}). Some terms came to the top of the frequency lists due to being mentioned in automated tweets, such as \emph{lastfm artists} and posts from \emph{foursquare}. 

Although much Twitter and internet-specific language was filtered using the stopword created using a sample of a Twitter Decahose, those with smaller frequencies still make it to the user terms. Thus, using three annotators, we manually exclude terms which would likely lead to poor bridges, including terms which may be too general (\emph{country}, \emph{things}), too ambiguous (\emph{analysis}, \emph{system}), or Twitter-specific (\emph{hashtag}), or too nonsensical (\emph{t20wc2014 fb}). As we describe below, we also use crowdsourcing to further clean this list of potential user interests during the survey construction. 

However, we do not restrict ourselves to the interests of the user, but utilize their network to link the user to other countries. We collect the user's social network by getting the user's $5,000$ friends and followers, and for those which are both friends and followers, their profiles and $3,200$ of their most recent tweets. For such ``reciprocal'' friends, we attempt to discern which country their location string belongs to using Yahoo PlaceMaker\footnote{\url{https://developer.yahoo.com/boss/geo/}}. Similarly, we extract the mentions of locations in various countries from the tweets these users have produced.

\subsection{Country Data}

Now that we've collected information on the user's potential interests and network, we turn to the countries. We use five resources to collect facts which are potentially interesting to the user. 

\textbf{Wikipedia} (\url{https://www.wikipedia.org/}) is a volunteer-based free encyclopedia which has been used as a resource for numerous studies. Its pages on countries include a summary of their geographic, economic, and political attributes, as well as cultural remarks. We collected data for 240 countries, with the text, stripped of HTML and other template content, having an average length of $3,363$ words. 

\textbf{Wikitravel} (\url{http://www.wikitravel.org/}) is another volunteer-based free encyclopedia which is designed for travelers to various destinations. Its articles explain the main geographical attributes of the location, how to travel to and from it, and what to see, do, and eat, as well as recommendations on where to stay. Whereas some of this information may overlap with Wikipedia, these pages provide an ``on the ground'' view of the location, with peculiarities of local culture, cuisine, and focus explicitly on features deemed to be interesting to a visitor. We collected Wikitravel articles for $177$ countries, with an average length of $3,929$ words. 

\textbf{Famous people} (\url{http://www.worldatlas.com/}), collected from the World Atlas site, are manually compiled lists of noted individuals from various countries, including politicians, artists, sportsmen, writers, and scientists. For each person, we attempt to search Wikipedia for a page on that person. Out of several possible pages returned by the Wikipedia API\footnote{\url{https://www.mediawiki.org/wiki/API:Main\_page}} we collect the number of page views each of these pages has gathered using Wikipedia Article Traffic Statistics website\footnote{\url{http://stats.grok.se/}} for 6 most recent months. The page which gets the most hits is picked. Finally, we get the abstracts for each page by parsing the page's HTML. The final collection consists of $178$ countries with an average of $135$ persons per country (with an exception of the United States, which had individual pages for each state, resulting in $6,398$ persons for the whole country). 

\textbf{Interesting facts} (\url{http://www.sciencekids.co.nz/}) about the countries were scraped from an educational website. The facts include geographical attributes of the country, its historical and cultural highlights, as well as local dishes and sports games. Select $52$ countries are covered by the website, with an average of $17$ facts per country.

\textbf{Web search} (\url{https://datamarket.azure.com/dataset/bing/search}), unlike the previous four sources, was performed with the user in mind, with the query consisting of a user interest (as gathered in the previous section) and the country name. The results then arguably represent webpages which link the two. Out of $4,158$ (\emph{user}, \emph{country}, \emph{interest}) triples (see User Survey Section for description of the users), $4,101$ gave at least one result using the Bing API\footnote{\url{http://www.bing.com/toolbox/bingsearchapi}}. In the next section we describe the cleaning procedures applied to these web search results.

\subsection{Building the Bridges}

Using the above data, we now are able build several kinds of bridges to address our research questions.

\subsubsection{Matching interests}

The country data we describe in the previous section contains hundreds of potential bits of information (sentences and paragraphs) which we may present to the user. To personalize this selection, we guide the selection of facts using the interests found in the user's tweets and profile. We begin the search process with the interests which are most frequently mentioned by the user. Among the retrieved candidate sentences (in the case of Wikipedia) or paragraphs (in the case of Wikitravel), we select those in which the interest word or phrase appears the earliest. Among the famous people, we select those with most Wikipedia page views, as a proxy to popularity.

The selection of web search results is more involved, as the Bing search results comprise a title, description, and URL. We score the results using the following equation:

\vspace{-0.4cm}
\begin{equation*}
\vspace{-0.2cm}
score = \alpha ~(t_c + t_i) + \beta ~(d_c + d_i) - \frac{rank}{\gamma}
\end{equation*}

where $t_c$, $t_i$, $d_c$, and $d_i$ are binary variables that indicate whether the (t)itle or (d)escription contain the (c)ountry or (i)nterest with weighting parameters of $\alpha = 30$, $\beta = 20$ and $\gamma = 10$. We filter the top $5$ results using an empirically determined threshold of $score$ $>$ $50$ as this requires that (i) at least one occurrence is in the title, and (ii) both occurrences are found in either the title or abstract. To break ties, use the one with the lowest rank.

In summary, we generate four personalized factual bridges using Wikipedia, Wikitravel, famous people, and web search. As a non-generalized alternative we use the interesting facts (due to being too few for a retrieval procedure similar to other sources).

\subsubsection{Utilizing User Network}

As we discussed earlier, social network has been used in recommendation before \cite{de2012chatter,konstas2009social,walter2008model}, but we propose to use it as explicit bridges to recommended items. Thus, we create two kinds of personalized bridges using the social network of the user. The first shows the mutual followers which are from a country, as identified by their location string. It is our intuition that the location of those connections which were made on the basis of interest outside the geographic constraints may be unknown to the user until pointed out. 

Second, we examine the tweets of the reciprocal followers (not necessarily from any particular country) which mention the country of interest. For example, they may mention news events (\emph{``Arizona wildfire near Yarnell kills 19 firefighters''}) and other facts (\emph{``70 years ago today 150,000 Allied troops landed in Normandy''}), but others express personal views (\emph{``What an amazing goal by Messi''}) or chatter with other Twitter users about the countries (\emph{``@handle Hope you’re having a great time. Try Sagardi or Taberna del la Reina''}). These highly personal views of the countries, as well as news and facts other users choose to retweet, may be more interesting to our users. This is not only because their social neighborhood is likely to share common interests, but also due to the simple fact that they were expressed by someone they know.


\section{User Study}
\label{sec:evaluation}

\begin{figure}
\begin{center}
\includegraphics[width=0.49\textwidth]{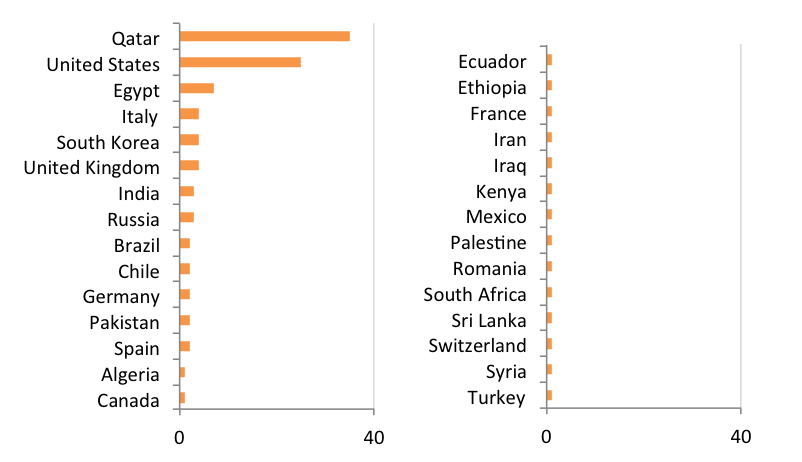}
\caption{Countries shown in descending order of the number of users in our study}
\label{fig:usercountries}
\end{center}
\end{figure}

We empirically test the proposed bridges via a user study. The $69$ participants were recruited via social media (Twitter and Facebook), with their Twitter handles and home countries collected. We defined a ``home'' country as that in which a participant has lived for some time, and we do not try to bridge the user to these countries. Figure~\ref{fig:usercountries} shows the countries from which these users have been recruited. The top two are Qatar and United States, with European and Asian countries following, making for a geographically and culturally diverse group. Using the Twitter handles, we proceeded to collect the participants' Twitter profile, tweets, and social network information, as described in the Twitter User Data Section. After processing this information and generating a candidate interest lists, we attempt to build a bridge between each user and each country. 

\begin{figure}[t]
\begin{center}
\includegraphics[width=0.30\textwidth]{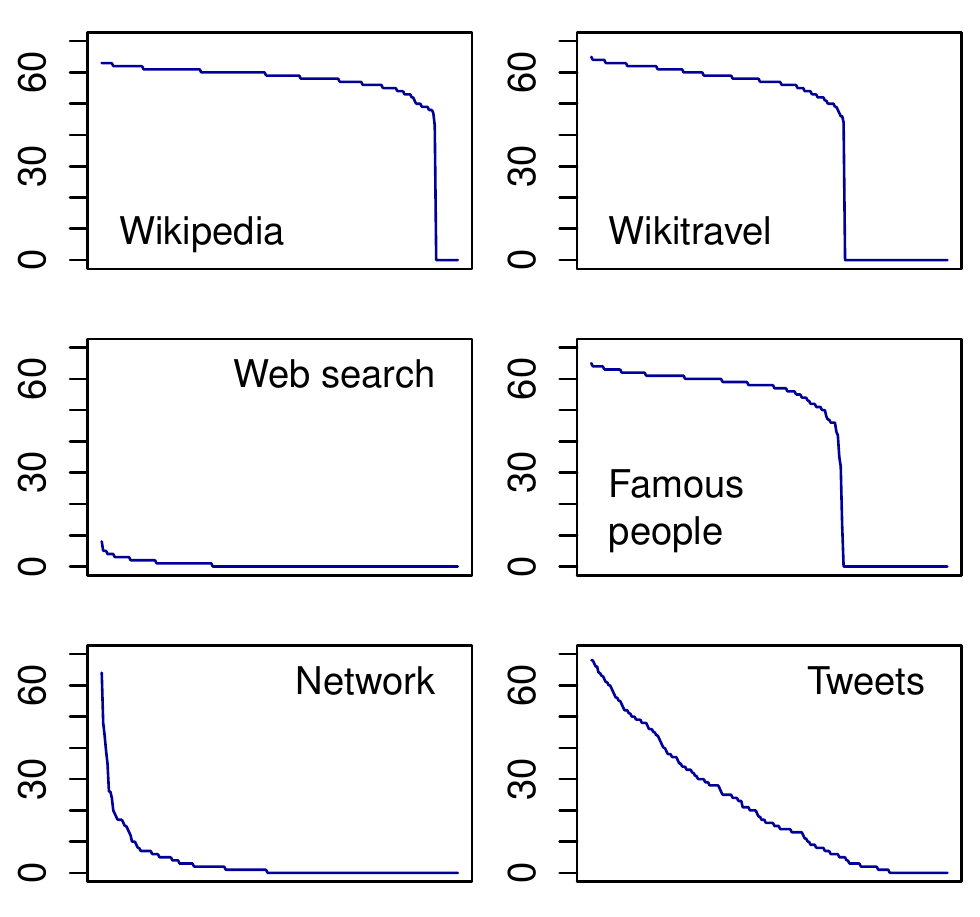}
\caption{Number of bridges generated for each country (x axis), with maximum of 69 user bridges}
\label{fig:bridgestats}
\end{center}
\end{figure}

Figure~\ref{fig:bridgestats} shows the coverage of the bridges for the six bridge types. The x-axis corresponds to countries, omitted for readability, sorted by the number of users with bridges to the country, and the y-axis shows the number of users which were able to be bridged to that country. The best coverage is provided by Wikipedia, Wikitravel, and the list of famous people. Web search resulted in very few bridges, due to our high quality threshold.

\begin{figure}[h]
\begin{center}
\includegraphics[width=0.30\textwidth]{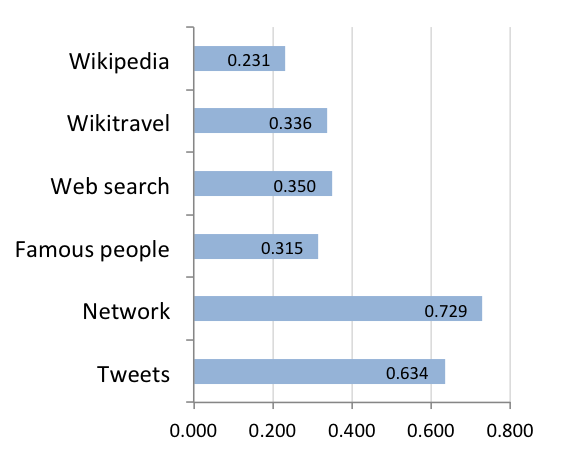}
\caption{Pearson correlation of country Wikipedia page views and the number of bridges generated}
\label{fig:bridgeviewscor}
\end{center}
\end{figure}

Intuitively, one would expect it to be easier to build bridges to the more well-known countries. Indeed, Figure~\ref{fig:bridgeviewscor} shows the Pearson correlations between the page views of Wikipedia page for each country with the number of bridges built for it. In all cases, the correlation is positive, but it is the greatest for bridges which utilize the Twitter social network and its content, underscoring the skewed distribution of society's attention \cite{zuckerman04nieman}. Although outside resources like Wikipedia and Wikitravel provide thorough coverage of most countries, and especially those which are well-known, social network and their tweets sometimes provide an alternative bridge to lesser-known countries such as Macedonia, Georgia, and Dominica. 

\begin{figure*}[t]
\begin{center}
\begin{subfigure}{0.40\textwidth}
  \centering
  \includegraphics[width=\textwidth]{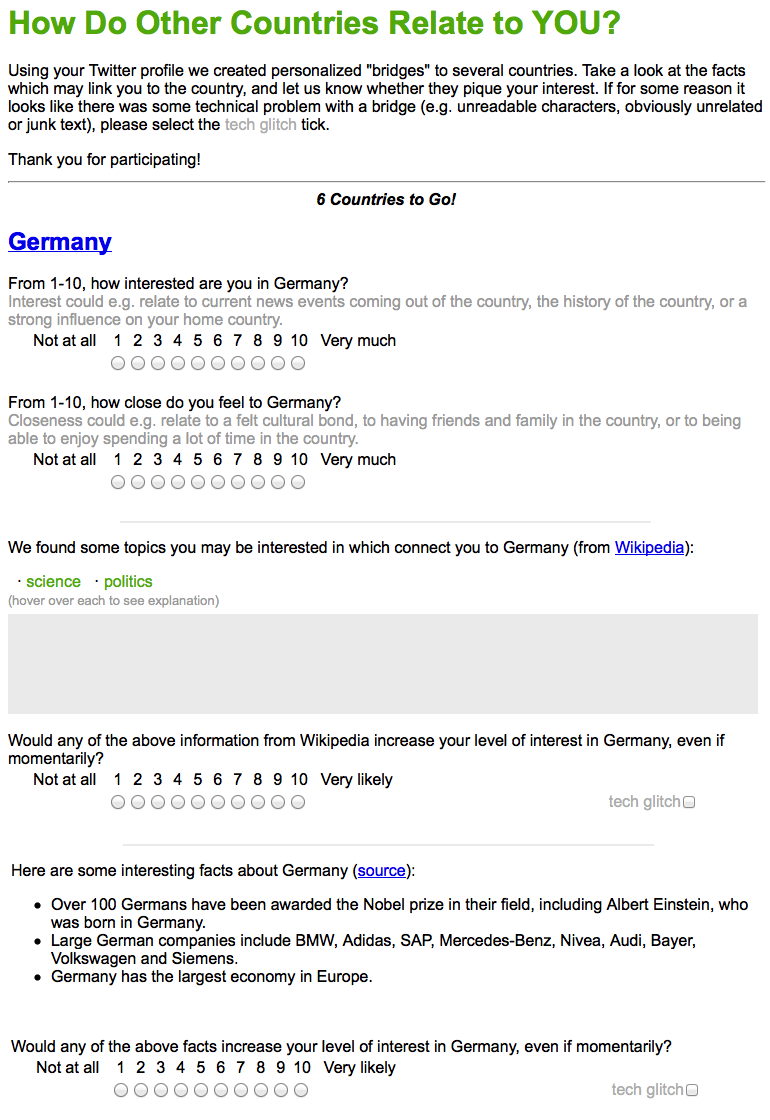}
\end{subfigure}
\hspace{1cm}
\begin{subfigure}{0.40\textwidth}
  \centering
  \includegraphics[width=\textwidth]{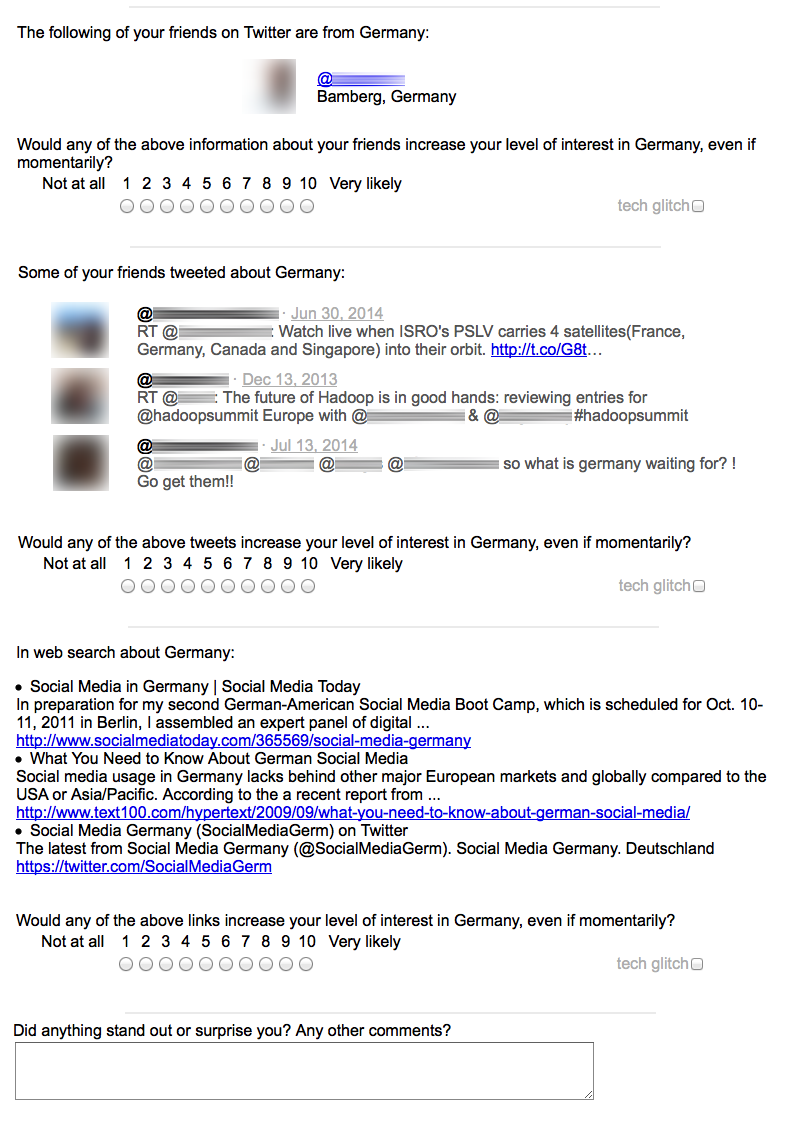}
\end{subfigure}
\caption{Screenshot of an example survey page}
\label{fig:screenshot}
\end{center}
\end{figure*}

Out of the generated bridges, we now need to select those which the user will annotate in our survey. To control for the popularity of the country, we classify the countries into well-known and little-known countries using the Wikipedia page views as a guide, taking the top third of the countries as well-known. We then select three well-known and four little-known countries for each user. We guide this decision by the number of bridges we were able to generate for a country, choosing those with most bridges first, randomly breaking ties (when countries have the same number of bridges). However, for each user-country combination, each kind of bridge has several candidates, and we use a crowdsourcing platform to ensure high quality of the selected bridges, as we describe next.

\subsection{Quality control}

We begin the selection of bridges by first ensuring that the words and phrases extracted from the Twitter profiles and tweets indeed represent a potential interest of the user. For this, we utilize the crowdsourcing platform CrowdFlower\footnote{\url{http://www.crowdflower.com/}}. For each $(\mbox{user}, \mbox{interest})$ pair, we show the labeler the Twitter profile including username, bio, location, and a selection of tweets which may mention this interest. We then ask whether the phrase represents a relevant interest of the user. We explicitly advise the users to remove non-sensical and overly generic candidates. The quality was ensured using a set of 28 ``gold-standard'' pre-labeled questions designed to be obvious to a labeler who has carefully read the instructions. The labeler had to answer correctly 4 out of 5 such questions before their responses were deemed trustworthy during the first task, and to continue answering correctly to one such question for consecutive tasks. Each labeling task consisted of 10 questions, worth 8 US cents. Finally, each of the 776 $(\mbox{user}, \mbox{interest})$ pairs was labeled by 3 independent labelers. The task proved to be fairly easy, with $85.49$\% inter-annotator agreement (in label overlap). Using these labels, we were able to disregard $12.6$\% of the potential interests.

Once the interests have been determined to be relevant to the user, we consider the various information which may serve as a bridge to a country. In a second CrowdFlower task we show the user a fact snippet coming from either Wikipedia, Wikitravel, famous people descriptions, or web searches, taking up to 6 candidate facts from each kind of bridge. We then ask the labeler whether the fact is closely related to the interest. Here, we aim to discard those snippets which are only tangentially or erroneously associated with the interest. As in the previous task, we use a set of 30 gold-standard questions for quality control, and collect 3 labels per $(\mbox{interest}, \mbox{fact})$ pair. This task has proven to be slightly more difficult than the previous, with an inter-annotator agreement of $82.9$\%. In total, $2,565$ $(\mbox{interest}, \mbox{fact})$ pairs were annotated, with $47.3$\% judged as irrelevant to the interest.
\\

\subsection{Survey}

Using the above high-quality set of bridges, we generated personalized surveys for each $(\mbox{user},\mbox{country})$ pair, with an example shown in Figure~\ref{fig:screenshot}. In the interest of space, the snippets were implemented to appear when user hovered the mouse over an interest, and links to further information (including the sources of the information) were included. We first assess the extent to which a user is already interested in a country, and how close they personally feel toward it. A status message at the top and bottom of the page indicated the number of countries left to go. At the bottom, a free-text field allowed participants to make remarks as they fill the survey. 

Additionally, two more pages were shown to the participants -- a welcome page which explained the goal of this survey and collected demographics, including age range, gender, and education, and a closing feedback page. The feedback allowed users to enter free text about their experience, as well as discuss privacy concerns they may have, and other information which may interest them about a country. 

The survey participants were invited to take the survey via email which they have submitted during the solicitation stage. Out of $67$ invited (2 were excluded due to near-empty Twitter accounts), $40$ have taken the survey, with a total of $253$ individual $(user, country)$ surveys submitted. Out of the participants, $29$ were male and $11$ female, with a median age between $31$ and $35$, and a high level of education, with majority having at least a Master's degree. 

Perhaps not surprisingly, our respondents showed more initial interest in well-known countries with an average score of $6.2$ out of $10$, compared to $4.3$ for little-known ones. We see the same trend for the personal closeness, with $5.2$ for well-known and $3.1$ little-known countries. Note that the interest scores were a whole point higher than the personal closeness one, indicating that participants were interested in countries beyond the personal connection. 

\begin{figure}[t]
\begin{center}
\includegraphics[width=0.45\textwidth]{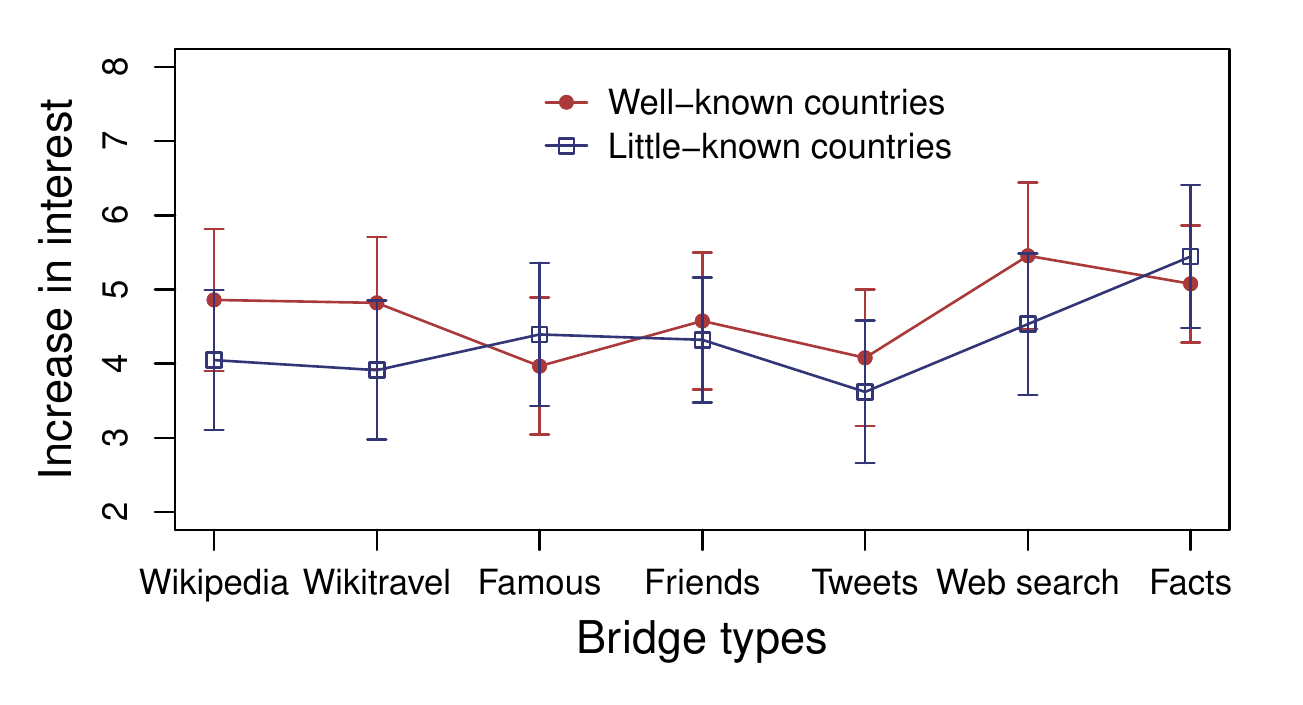}
\caption{Interest increase for each kind of bridge for well-known and little-known countries, with 95\% confidence intervals.}
\label{fig:interestincrease}
\end{center}
\end{figure}

\begin{figure}[t]
\begin{center}
\includegraphics[width=0.35\textwidth]{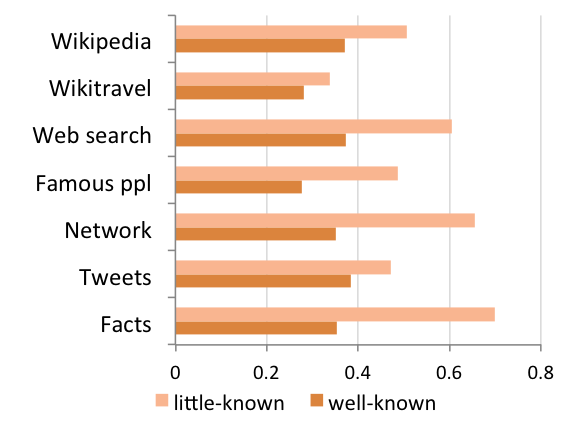}
\caption{Pearson correlation between the initial interest in the country and the interest increase by each kind of bridge.}
\label{fig:correlinterest}
\end{center}
\end{figure}

We further examine the increase in interest for each bridge type in Figure~\ref{fig:interestincrease} for the two groups of countries (with 95\% confidence intervals). First note that the average increase in interest ranges between $3.5$ and almost $6.0$ out of $10$, indicating the validity of some of our bridging attempts. The difference between the Wikipedia and Wikitravel interest scores is significant at the $p$-values of $0.088$ and $0.073$ in the favor of well-known countries, whereas the others have even less significant difference. 
Note that the manually-created ``interesting facts'' method has resulted in the best interest improvement, suggesting the automated methods still need some improvement. Furthermore, we find a marked difference in the way the participants' initial interest affected their subsequent interest in the bridges. Figure~\ref{fig:correlinterest} shows that the initial interest expressed in the country predicts the interest increase for little-known countries much more than their better known neighbors. This once more underscores the difficulty of connecting people to un-common information, especially if they have little interest in it to begin with. The feedback in our subsequent interviews further emphasized the importance of matching known interests of the user when introducing new information.

The bridges which were judged to be interesting included hobbies like \emph{robotics} and \emph{biking}, political interests like \emph{revolution} and \emph{politics}, and professional ones like \emph{social media} and \emph{science}. A selection of bridges built using Wikipedia, Wikitravel, and famous people is shown in Table~\ref{tbl:examplebridges}. Wikipedia tends to provide more historic perspective, such as a note on the American revolution, whereas Wikitravel provides a more personal on-the-ground information about travel routes and activities. Famous people snippets often provide both a historic perspective on the countries as well as a more international angle, as many cited people have traveled in or interacted with several countries, such as the Korean-born American football player shown for interest \emph{pittsburgh}.

\begin{table*}
\caption{Example bridges built using Wikipedia, Wikitravel, and Famous people as sources.}
\label{tbl:examplebridges}
{\centering\scriptsize
\begin{tabular}{lllp{12.5cm}}
\toprule
\textbf{Interest} & \textbf{Country} & \textbf{Source} & \textbf{Quote}\\\midrule
\emph{robotics} & South Korea & Wikipedia & Robotics are also incorporated in the entertainment sector as well; the Korean Robot Game Festival has been held every year since 2004 to promote science and robot technology.\\
\emph{revolution} & United States & Wikipedia & The American Revolution was the first successful colonial war of independence against a European power.\\
\emph{arabic} & Bahrain & Wikipedia & Arabic calligraphy grew in popularity as the Bahraini government was an active patron in Islamic art, culminating in the establishment of an Islamic museum, Beit Al Quran.\\
\emph{damascus} & Jordan & Wikitravel & Long distance taxis used to operate the route from Damascus to Amman. However, due to the current status in Syria you are unlikely to be able to travel from Damascus to Amman via taxi, as anyone out in the open is at risk of being shot by either government-controlled powers or the rebels.\\
\emph{bike} & Bhutan & Wikitravel & There is no better way to experience a place than by bike. Bhutan's expansive wilderness, small sleepy cities, rural farmland, tiny roads, majestic views, and everything else is that much sweeter on a bike. Biking isn't easy though, being in the Himalayas it requires long climbs at higher altitudes than most travelers are accustom to, followed by long descents down roads rarely wide enough for two cars. \\
\emph{pittsburgh} & South Korea & Famous people & Hines Ward (born Mar 8, 1976) is a former American football wide receiver and current NBC studio analyst who played fourteen seasons for the Pittsburgh Steelers of the National Football League (NFL). Born in Seoul, South Korea to a Korean mother and an African American father, Ward grew up in the Atlanta area. [...] \\
\emph{tanks} & Russia & Famous people & Mikhail Koshkin (Dec 3, 1898 - Sep 26, 1940) was a Soviet tank designer, chief designer of the famous T-34 medium tank. The T-34 was the most produced tank of World War II. He started out in life as a candy maker, but then studied engineering. [...] \\
\bottomrule
\end{tabular}}
\end{table*}

As mentioned earlier, the users were able to let us know if something did not seem right with the bridge by selecting a ``tech glitch'' tick. User network's tweets mentioning the countries received the most ticks at $8$ (out of $214$ instances) and $21$ comments remarking on the poor matching of tweets to the countries. There are several reasons for these mismatches: the ambiguity of online language with abbreviations like ``CA'' (which may match to Canada or California), different languages and alphabets (our dataset included a large amount of Arabic, as well as Korean, Turkish, German, etc.), and ambiguous word choice (like ``new york steak''). In Discussion Section we propose some possible technical solutions to these problems.

This exploratory survey gave us a first glimpse into our ability to provide personalized bridges to countries, finding an interplay of user's initial knowledge about the country, their interests, and the kind of informational snippets we provide. However, the long-form feedback pointed to a complex roles of personal interests, personal versus professional emphasis of the Twitter account, and the information presentation. Next, we interview $11$ participants (who come from all over the world) to learn more.


\section{Interviews}
\label{sec:interview}

In order to get a more elaborate feedback on the bridges our study participants have witnessed, we conducted $11$ interviews of select users. Six male and $5$ female users were from one of the following countries: United States, South Korea, Algeria, India, Turkey, Italy, and Bulgaria, spanning North America, Europe, Middle East, and Asia. The interviews were performed both online (using voice or video chat) and in person. The interviews lasted between $15$ and $45$ minutes, on average lasting about $30$ minutes, and were driven using a slideshow and a set of pre-determined questions, which were organized as follows: 

\begin{enumerate}[itemsep=0mm]
\item The interviews started with an explanation of the goal of the study and a discussion of its benefits and possible difficulties. 
\item We then solicited feedback about the bridges built for the study, overall impression and bridges which were specifically remembered by the participants. 
\item We then showed 3-5 additional bridges hand-picked for the participant to explore potential directions.
\item We finally asked the user exploratory questions about additional information they may find interesting as a bridge, and other sources of information (such as Facebook) which may be used to mine their interests.
\end{enumerate}

All of the participants have found the goal of the project interesting and potentially beneficial. However, participant 7 was encouraging us to phrase our goal to make sure we are not suggesting \emph{new interests} -- in which one is highly unlikely to succeed -- but instead \emph{exposing} people to new information they would not encounter in their daily lives. Some stated that they already found ways of enriching their news interests by following Twitter accounts of informational resources like \emph{@Stratfor}\footnote{\url{http://www.stratfor.com/}} (participant 9). None, of course, were personalized, though.

The interviewees provided a set of valuable feedback on the quality of the bridges. The following aspects of the bridges were particularly emphasized:

\begin{itemize}
\setlength\itemsep{0cm}
\item \textbf{Relevance}. Once the interests have been determined, they must be central to the information (facts and tweets) presented. For most, it was not enough that their contacts were mentioning some country in their tweets -- that is, network alone does not substitute an interest model. 
\item \textbf{Recency}. Both extracted interests and the accompanying facts should be recent and up-to-date. On the side of interest extraction, more weight should be given to keywords mentioned more recently. On the side of facts, participant 1 emphasized that he is interested only in the current events and participant 6 that the statistics on the countries should be up to date (especially about fast-changing phenomena like internet use).  
\item \textbf{Novelty}.  Not only the information must be recent, but also it needs to be sufficiently different from the daily ``chatter'', such as current popular news or ongoing professional events which may already be known to the user (participants 4 \& 6).
\item \textbf{Personal closeness}. Whether a contact is a personal or professional one may make a difference, with updates from personal contacts being more welcome, while personal updates from professional contacts feeling ``too private'' (participant 11). Both spheres also have degrees of importance, though, illustrated when participant 8 noticed a tweet from her close supervisor, and admitted that she would pay much greater attention to it than to other professional contacts.
\item \textbf{Language}. The language of the bridge may convey not only the sphere of the interest, but would also introduce a greater variety of sources (participants 2 \& 8), such as Wikipedia articles in other languages.
\item \textbf{Media}. Several participants (2, 3, 5) were suggesting that photos, graphics or cartoons would be eye-catching and may convey information quicker than text.
\item \textbf{Brevity.} It would be the best, said participant 9, for the bridge to be as small as a tweet. A number of survey responders also commented on the length of some of the bridges to be prohibitive.
\item \textbf{Emotion.} Finally, participant 4 expressed a greater interest in provocative, emotional, ``flamish'' (as in trying to start a ``flame war'') content, which, if coming from trusted sources, would indicate a topic they really cared about.
\end{itemize}

Many responders emphasized the highly individual tastes which would affect whether one is open to exposure to new information. One have mentioned her friends who do not travel internationally and have few friends abroad, who have ``entirely different view of the world'' and who may be less interested in other countries. Sensitivity to the world views of individuals around the world is also needed to properly understand the particular meaning other countries may have for them. Participant 2, for instance, cited a tense relationship between India and Pakistan, such that travel and communication between the two neighboring countries is limited. In this case, she found personal links to the other country especially intriguing, seeking to better understand commonalities between the communities. On the other side, participant 4 expressed indifference towards the more familiar European neighbors, and suggested that a friend's presence in a more exotic place would be more captivating. Cultural sensitivity, therefore, may be needed to both understand whether a bridge is to a common and ``boring'' or a new or even excitingly-``forbidden'' place.


\section{Discussion}
\label{sec:discussion}

Barring the technical challenges (which we address below), the extent of the efficacy of our approach depends on the content users are willing to share on Twitter. A majority of our interviewees used Twitter for professional means, as well as personal (although Participant 2 had three profiles, each geared to different spheres of their life). A multi-purpose Twitter profile then has social connections which may reflect one or several facets of the user. Such relationships were defined by Barnes \cite{barnes1979network} as a property of \emph{multiplexity} in network ties -- those which serve more than one purpose or social activity. Recent work on detecting professional versus personal closeness in enterprise social networks by Wu {\em et al.} \cite{wu2010detecting} illustrates some features which may distinguish one from another, including sharing professional interests, frequency of communication and the activity level of each individual user. Classifying user contacts into professional and personal (and possibly other categories), as well as establishing the closeness of the relationship \cite{xie2012friend}, may help in isolating the neighborhood which may provide more appealing bridges. 

Ambiguity was not only an issue for social connection selection, but also for the extraction and matching of user interests. Here, alternative sources of information, as well as additional NLP tools may be employed. First, additional information is often provided in links shared through tweets, including pictures from Instagram, locations from Foursquare, and even public Facebook posts. A user's webpage may provide an alternative window into their online persona (which may also be of a professional nature). Favorited tweets may show the user's long-term interests. Finally, clustering and segmenting the contacts of the user into ``interest groups'' may help identify related concepts and enrich our understanding of the context of a user's posts. Furthermore, at matching time, identifying good candidate facts, using an \emph{aboutness} score \cite{moens2006measuring}, which employs coreference and reference resolution, would result in a selection of more relevant texts. Finally, expanding the concepts found both in user interest models and other texts using knowledge bases (such as Freebase\footnote{\url{http://www.freebase.com/}}) could provide more semantically rich connections.

In an attempt to gauge the sentiment about privacy issues, we asked our interviewees about the mining of their social media profile data, and especially that on other websites, including Facebook. Although all felt comfortable with Twitter being used for studies, most have expressed concerns about Facebook. Participant 9 suggested to give the user control over which parts of the account can be used for mining, while Participants 3, 6, 8 would only allow an application access to their Facebook data if they knew its legitimacy (if it came directly from this research group, for instance). Participant 7 even indicated that they would not welcome ``bridging'' suggestions on Facebook (but would on Twitter). These findings emphasize the need for clear communication between the application developer and its users regarding data use concerns.

The aim of this project is to break through the ``filter bubble'' surrounding each of us by taking advantage of more tenuous connections, much like in Granovetter's ``weak ties'' \cite{ granovetter1973strength}, to cross group boundaries and compel the user to relate to new information. For example, we envision a use of personalized bridging in news websites, where along with an article additional information relating the reader to its events and entities would catch and keep user's attention. But to truly overcome the news coverage biases \cite{zuckerman04nieman}, the news presentation and dissemination should be adjusted. As future work we will create a news recommender Twitter account, which will generate personalized news links to its followers (thus making it an opt-in, low-hassle engagement for the user). The exploratory nature of this project makes it particularly important to give a user a chance to express their willingness to participate by opting in to receive the messages. The same strategy was followed in the user study of this paper, providing us with interested and attentive individuals to develop the system. Eventually, though, our aim is to generalize this system to be able to improve the visibility of any long-tail content, including news stories, societal issues, and far-away places. For example, the same algorithms may be used in promoting other content, such as charitable causes in little-known countries. 

Many questions remain, though. How long will a bridge ``hold''? How can we evaluate such bridges on a massive scale? Will users be patient with such unusual recommendations in the long term? Our future work in building a real-time personalized bridge generator will hopefully answer some of these questions.

\section{Conclusions}

In this work we attempt to build personalized ``bridges'' to lesser-known content. Using open-source online resources, as well as user social network, we attempt to interest users in well-known as well as lesser-known countries. On average, our recommendation resulted in 4 to 5/10 interest increase, as measured using $253$ user/country survey responses. We show that the lesser-known countries are not only more challenging to find bridges for, but also that the user's initial familiarity of it dramatically influences  the effect of the bridge. During the interviewing effort of $11$ Twitter users we find a complex \emph{multiplexity} in the kinds of relationships fostered there, and provide a set of requirements effective bridges must meet. 

The results of this study will inform the creation of a system for relating little-known content to social media users, in attempt to break the barriers of technology and culture, and give a voice to the voiceless.

%
%
%
%
%
\balance


\bibliographystyle{acm-sigchi}
\bibliography{relatability}
\end{document}